\documentclass[%
 reprint,
 superscriptaddress,
 amsmath,amssymb,
 aps,
 prb,
]{revtex4-2}
\usepackage[utf8]{inputenc}
\usepackage[T1]{fontenc}
\usepackage{graphicx}
\usepackage{dcolumn}
\usepackage{bm}
\usepackage{svg}
\usepackage[english]{babel}
\usepackage[hidelinks]{hyperref}

\begin{document}

\title{Approaching the Fundamental Limit of Single-Shot Qubit Frequency Tracking with an Adiabatic Tangentially-Modulated Pulse}

\author{Itamar Oren}
\email{i.oren@student.unsw.edu.au}
\affiliation{%
 School of Electrical Engineering and Telecommunications, The University of New South Wales, Sydney, NSW 2052, Australia
}%

\author{Luke I. Dyer}%
\email{luke.dyer@unsw.edu.au}
\affiliation{%
 School of Electrical Engineering and Telecommunications, The University of New South Wales, Sydney, NSW 2052, Australia
}%

\author{Calum Holker}
\affiliation{%
    School of Electrical Engineering and Telecommunications, The University of New South Wales, Sydney, NSW 2052, Australia
    }%
\affiliation{%
    Diraq Pty Ltd, Sydney, NSW, Australia}%

\author{Gerardo A. Paz-Silva}
\affiliation{%
    Diraq Pty Ltd, Sydney, NSW, Australia}%

\author{Chih Hwan Yang}
\email{henry.yang@unsw.edu.au}
\affiliation{%
    School of Electrical Engineering and Telecommunications, The University of New South Wales, Sydney, NSW 2052, Australia
    }%
\affiliation{%
    Diraq Pty Ltd, Sydney, NSW, Australia}

\date{\today}

\begin{abstract}
Understanding and mitigating noise in two-level quantum systems is essential for achieving high-fidelity qubit control. Conventional frequency tracking techniques, such as Ramsey interferometry, are fundamentally limited by trade-offs between sensitivity, bandwidth, and dynamic range. Here we introduce the adiabatic tangentially-modulated (ATM) pulse, a pulse derived from quantum adiabatic theory that maps qubit detuning onto a sigmoidal, near-binary response. Using numerical simulations supported by analytical modeling, we show that pulse sensitivity and detuning range can be independently engineered through simple design parameters. We derive scaling relations governing these quantities and demonstrate their agreement with simulation. A single-shot ATM measurement achieves sensitivity comparable to that obtained from multi-shot Ramsey averaging, enabling tracking of substantially higher frequency noise components with a lower closed-loop white-noise floor. In addition, this pulse exhibits strong robustness to amplitude fluctuations compared with binary response pulses derived from the Shinnar-Le Roux formalism. Together, these properties establish the ATM pulse as a promising approach for robust qubit frequency tracking.

\end{abstract}

\maketitle

\section{\label{sec:Introduction}Introduction}

Precise knowledge and control of the resonant frequency is vital to qubit operations \cite{Takeda_2018}. Drift and fluctuations in that frequency, driven by a multitude of factors such as magnetic-field noise, charge noise, and device instabilities, degrade gate and readout fidelity unless actively tracked and compensated \cite{Proctor_2020, PhysRevApplied.10.044017, Shehata_2023, Capannelli_2025, Hensen_2020, Yang_2019, Shulman_2014}. A number of protocols exist for frequency sensing and stabilization, the most common being Ramsey-style interferometry \cite{PhysRev.78.695, RevModPhys.89.035002, steinacker_industry-compatible_2025, vepsalainen_improving_2022, nakajima_coherence_2020, dumoulin_stuyck_silicon_2024}, which can provide real-time feedback but remains fundamentally limited to compensating noise only below a certain frequency \cite{Hecht_2025, rojasarias2024originsnoisezeemansplitting, nakajima_coherence_2020, dumoulin_stuyck_silicon_2024, yoneda_quantum-dot_2018}, regardless of the number of shots used. Further, Ramsey interferometry has trade-offs between speed (number of shots), sensitivity, and dynamic range due to underlying periodicity \cite{PhysRevApplied.19.064022} (explained further in Section \ref{sec:NoisePSD}). Alternative pulse design techniques developed for nuclear magnetic resonance can also generate sharp, near-binary frequency discrimination responses \cite{shinnar1989synthesis,grissom2014b1, Kessler1991}. These approaches provide a useful point of comparison for the frequency tracking protocol presented in this work.

In this work, we introduce the Adiabatic Tangentially-Modulated (ATM) pulse, a single-shot driven protocol that uses adiabatic evolution similar to other frequency-modulated qubit control pulses \cite{Kuzmanovic2024HighFidelity,McCord2025Pareto,Kuzmanovic2025NeuralNetwork}. The ATM pulse converts the continuous detuning variable into a sigmoidal (near-binary) readout with controllable sensitivity and range. Here, the fundamental limit refers to an ideal single-shot binary discrimination step with unit contrast and a vanishing transition width. The ATM pulse uses three key features: a tangent-shaped amplitude rise, a piecewise-linear frequency chirp that maps detuning into an instantaneous eigenbasis, and a controlled tangent-shaped fall stage that determines the adiabatic to non-adiabatic crossover and therefore the system sensitivity. Together, these features create a pulse that offers improved frequency tracking performance relative to Ramsey-based methods while maintaining robustness to amplitude errors that would otherwise degrade alternative binary-response pulse schemes.

\section{\label{sec:ATM}Hamiltonian Formulation and Construction of the ATM Pulse}

The key resource is qubit driving using an amplitude- and frequency-modulated transverse field, effectively creating time-dependent Rabi and driving frequencies. Analogous to conventional Rabi driving, the qubit experiences a controllable static field along the $z$-axis together with a field oscillating along the $x$-axis. The laboratory frame Hamiltonian is written as
\begin{eqnarray} 
    \mathcal{H}_{\mathrm{lab}}(t) = \omega_0 S_z + 2\Omega(t)\cos\left(\omega t + \phi_D(t) \right) S_x , 
\end{eqnarray}
where the field amplitude $\Omega(t)$ and drive phase $\phi_D(t)$ may both be shaped in time, giving full control over the instantaneous frequency and amplitude of the drive.

Transforming into a frame rotating at the carrier frequency $\omega$ and defining the detuning $\Delta = \omega_{0} - \omega$ (see Appendix \ref{sec:Rotating}), the Hamiltonian in the rotating frame is
\begin{eqnarray} \label{eq:RotatingHamiltonian}
    \mathcal{H}_{\mathrm{rot}}(t)
    & = & \Delta S_z \nonumber \\
    &   & + \Omega(t)\cos\big(\phi_D(t)\big)\, S_x \nonumber \\
    &   & + \Omega(t)\sin\big(\phi_D(t)\big)\, S_y .
\end{eqnarray}

For simplified notation, we can define $\Omega_x(t) =\Omega(t)\cos\big(\phi_D(t)\big)$ and $\Omega_y(t) =\Omega(t)\sin\big(\phi_D(t)\big)$. Further, if the phase is written as
\begin{eqnarray} 
    \phi_D(t) = \int_{0}^{t} \omega_D(\tau)\, d\tau \, , 
\end{eqnarray}
then $\omega_{D}(t)$ is the instantaneous frequency offset from the carrier $\omega$ and is fully tunable through external control.

\begin{figure}[b]
    \includesvg[width=\linewidth]{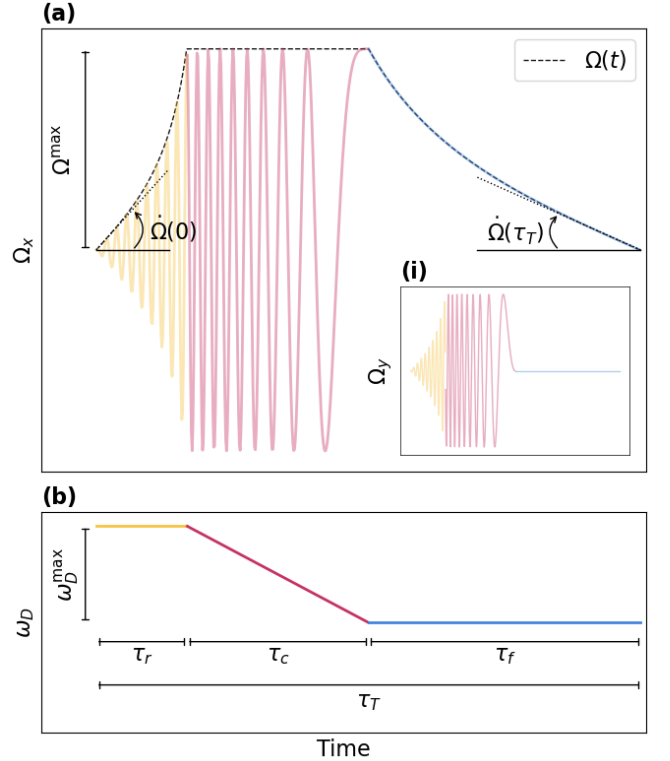}
    \caption{Construction of the adiabatic tangentially-modulated (ATM) pulse. (a) Amplitude modulation of the driving field, expressed as the Rabi frequency envelope $\Omega(t)$ [Eq. (\ref{eq:AM})]. The envelope consists of three segments: a tangent-shaped rise of duration $\tau_r$ with initial gradient $\dot{\Omega}(0)$, a constant-amplitude section of duration $\tau_c$ at maximum amplitude $\Omega^{\max}$, and a tangent-shaped fall of duration $\tau_f$ with final gradient $\dot{\Omega}(\tau_T)$. The full pulse length is $\tau_T = \tau_r + \tau_c + \tau_f$. Dashed lines illustrate the tangent gradients that set the boundary conditions of the rise and fall. The oscillatory trace shows the mixed envelope and driving signal for $\Omega_x$, while inset (i) shows the quadrature component $\Omega_y$. (b) Frequency modulation of the driving field, $\omega_D(t)$ defined by piecewise-linear chirp [Eq. (\ref{eq:FM})]. During the rise, the drive frequency is held constant at its maximum $\omega_D^{\max}$. Over the central interval $\tau_c$, the frequency chirps linearly from $\omega_D^{\max}$ to zero. In the $\tau_f$ segment, the drive remains fixed at zero. Together, the coupled amplitude and frequency modulations define the full ATM pulse profile used to achieve adiabatic state transfer.}
    \label{fig:ATMPulseParameters}
\end{figure}

The Adiabatic Tangentially-Modulated (ATM) pulse is constructed through the combined modulation of the driving amplitude and frequency illustrated in Figure~\ref{fig:ATMPulseParameters} and described below. The ATM pulse is separated into three piecewise segments, with the amplitude and drive-frequency offset continuous at their boundaries.

\subsection{\label{sec:AM} Amplitude Modulation}
The amplitude profile is composed of rise, constant-amplitude and fall segments, denoted $\tau_r$, $\tau_c$, and $\tau_f$, respectively (noting that $\tau_T$ is the sum of these three):
\begin{eqnarray} \label{eq:AM}
\Omega(t) =
\begin{cases}
    \frac{\dot{\Omega}(0)}{\beta_r} \tan(\beta_r t), & 0 \leq t < \tau_r, \\
    \Omega^{\mathrm{max}}, & \tau_r \leq t < \tau_r + \tau_c, \\
   -\frac{|\dot{\Omega}(\tau_T)|}{\beta_f} \tan\left[\beta_f (t-\tau_T)\right], 
   & \tau_r + \tau_c \leq t \leq \tau_T.
\end{cases}
\end{eqnarray}
Here, the parameters $\beta_r$ and $\beta_f$ are chosen so that $\Omega(t)$ is continuous across each boundary.  Tangent-shaped rise and fall sections were selected because they provide a smooth amplitude transition with finite slope and controlled curvature. Specifically, this form is directly motivated by the geometry of two-level dynamics, where the polar angle of the Hamiltonian on the Bloch sphere is related to the ratio of the Rabi amplitude and detuning through
\begin{eqnarray}
\theta(t)=\operatorname{atan2}\!\left(\Omega(t),\Delta\right),
  \qquad
  \tan\left(\theta(t)\right)=\frac{\Omega(t)}{\Delta},
\end{eqnarray}
where the ratio form applies away from $\Delta=0$. Thus a tangent-shaped $\Omega(t)$ provides a convenient form for smoothly varying $\theta(t)$ during the rise and fall segments. 

\subsection{Frequency Modulation \label{sec:FM}}
The instantaneous drive-frequency offset $\omega_D(t)$ is defined by a piecewise linear chirp profile,  
\begin{eqnarray} \label{eq:FM}
\omega_D(t) =
\begin{cases}
    \omega^{\max}_D, & 0 \leq t < \tau_r, \\
    \omega^{\max}_D - \dfrac{\omega^{\max}_D}{\tau_c}(t-\tau_r), 
    & \tau_r \leq t < \tau_r + \tau_c, \\
    0, & \tau_r + \tau_c \leq t \leq \tau_T,
\end{cases}
\end{eqnarray}
so that the chirp decreases linearly during the central interval $\tau_c$ and is zeroed out during the fall stage. The piecewise linear chirp maps the drive-frequency offset from an initial offset $\omega_D^{\max}$ to zero. This linear sweep establishes a controllable instantaneous eigenbasis trajectory that depends on the detuning $\Delta$, as explained in Section \ref{sec:Understanding}. \\

Notably, the sign of the chirp is conventional: reversing it mirrors the response about $\Delta=0$ without changing its width or range. The response simulations use the negative sweep specified in Appendix~\ref{sec:Pulse}.

\section{\label{sec:Understanding}Geometric Interpretation of the ATM Pulse}

\begin{figure}[b] 
    \includesvg[width=\linewidth]{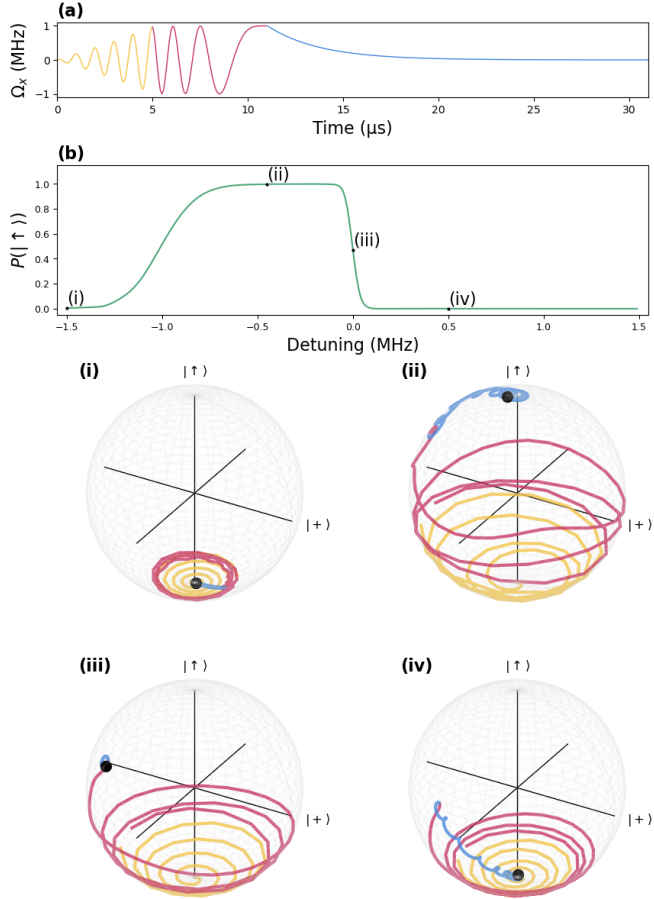}
    \caption{(a) Temporal profile of ATM pulse (see Appendix~\ref{sec:Pulse} for pulse parameters), divided into its three sequential segments: rise ($\tau_r$, yellow), constant-amplitude chirp ($\tau_c$, red), and fall ($\tau_f$, blue). (b) Resulting transition probability as a function of detuning, highlighting four key detuning points labeled (i–iv). (i–iv) Bloch sphere representations of the qubit state in the rotating frame at the detuning values indicated in (b). The color of each trajectory denotes the pulse segment during which the corresponding state evolution occurs, as defined in panel (a).}
    \label{fig:BLOCH}
\end{figure}

A clearer understanding of the ATM pulse arises from the geometric picture of the qubit evolution on the Bloch sphere. As shown in Figure~\ref{fig:BLOCH}, the rise and chirp segments prepare the system in a qubit state whose angle on the Bloch sphere depends strongly on the detuning. During the rise, the tangent-shaped amplitude profile ensures a smooth and controlled turn on of the drive amplitude. The constant-amplitude chirp then sweeps the drive-frequency offset, causing the qubit state to evolve in a detuning-dependent manner; this establishes a structured mapping between detuning and the position of the qubit state relative to the $xy$-plane.

The final tangential fall plays an important role in determining the qubit’s final state. For trajectories that enter this segment close to an instantaneous eigenstate, the fall is designed to promote adiabatic movement for states lying predominantly above or below the $xy$-plane, while the evolution becomes increasingly nonadiabatic near the end of the fall for small nonzero detunings (as discussed further in Section \ref{sec:DetuningOptimisation}). As a result:
\begin{itemize}
    \item For trajectories lying below the $xy$-plane, the adiabatic fall draws the state downward, yielding a final population close to the ground state (Figure \ref{fig:BLOCH}(i \& iv)).
    \item For trajectories lying close to the $xy$-plane -- which typically corresponds to $\Delta = 0$ -- the fall maintains the state near a $|-X\rangle$ orientation (Figure \ref{fig:BLOCH}(iii)).
    \item For trajectories lying above the $xy$-plane, the fall drives the system upward toward the excited state, producing a transition (Figure \ref{fig:BLOCH}(ii)).
\end{itemize}

This geometric perspective illustrates the creation of the excitation range and the sharp sensitivity edge observed in the ATM response. The chirp stage prepares the system into one of three qualitative state regions (below, on, or above the equator), and the fall segment maps these trajectories onto distinct final outcomes. A further drive-following description of this evolution and its adiabaticity is given in Appendix~\ref{sec:Adiabaticity}.

\section{\label{sec:DetuningOptimisation}Detuning Parameterization and Pulse Optimization}
\begin{figure}[b] 
    \includesvg[width=\linewidth]{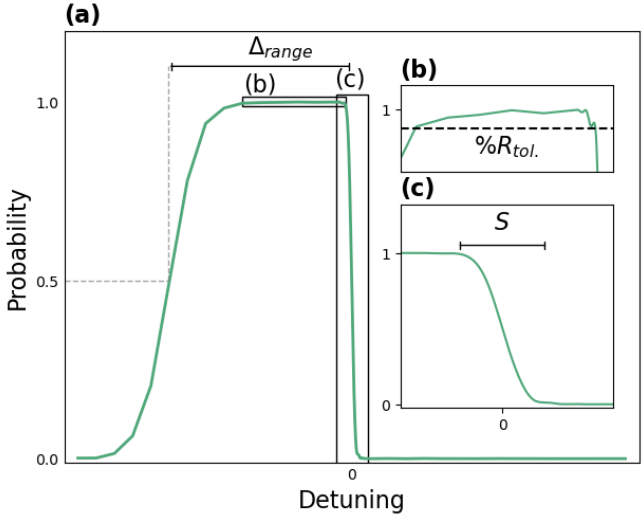}
    \caption{Detuning-dependent transition probability produced by the ATM pulse. (a) Probability $P = |\langle \uparrow|\psi(\tau_T, \Delta)\rangle|^2$ as a function of detuning $\Delta$. The system produces a nearly flat excitation plateau over the detuning window $\Delta_{\mathrm{range}}$. (b) Magnified view of the response near the upper flat section, highlighting the system's ripple. The dashed horizontal line marks a user-defined ripple tolerance threshold $\%R_{\mathrm{tol.}}$, which helps define the acceptable range across detunings. (c) The system sensitivity $S$, where the probability of transfer switches from an excited to unexcited qubit and thus to its selectivity in qubit resonance.}
    \label{fig:EvolvedProb}
\end{figure}

An optimized ATM pulse, initialized in $|\downarrow\rangle$ and evolved to the final state $|\psi(\tau_T)\rangle$, produces the detuning-dependent transition probability
\begin{eqnarray}
    P(\Delta) = \big|\langle \uparrow | \psi(\tau_T,\Delta) \rangle\big|^2 ,
\end{eqnarray}
shown in Figure \ref{fig:EvolvedProb}.

For optimized parameters, the transition probability exhibits a range of detuning with an excited state, followed by a sharp transition to the complementary ground state range. Thus, qubit states evolved under the ATM pulse map continuous detuning values onto two distinct and well-separated outcome classes, which may be interpreted operationally as binary readout states. 
The sharp edge of the resulting response enables the ATM pulse to function as a single-shot qubit frequency tracker. By monitoring whether the system ends in the excited or ground state, one can determine which side of the effective resonance condition the detuning lies on, provided that the tracker remains within its operational range. As a result, the design objective is to appropriately tune the pulse characteristics to achieve a desired sensitivity and operational range while maintaining low ripple across the plateau. These design parameters allow the ATM pulse to accommodate different experimental constraints, such as maximum drive power or time limitations, while preserving high-sensitivity binary readout of the qubit detuning.

\subsection{Optimizing for Sensitivity}
The Quantum Adiabatic Theorem states that if a time-dependent, nondegenerate quantum system is initially in the $n$-th eigenstate of $\mathcal{H}(t)$, then the state remains in the corresponding instantaneous eigenstate of $\mathcal{H}(t)$ at all times $t$ provided that $\mathcal{H}(t)$ evolves sufficiently slowly \cite{PhysRevA.80.012106}. For the ATM pulse, this condition formally corresponds to $\tau_T \rightarrow \infty$, which is not physically achievable. A widely used practical condition for adiabatic evolution is
\begin{eqnarray}
\frac{\hbar\left|\langle E_m(t) | \dot{ \mathcal{H}(t)}|E_n(t) \rangle \right|}{\left|E_m(t) - E_n(t)\right|^2}\ll 1, \quad m \neq n, \quad t \in [0, T],
\end{eqnarray}
where $E_{m,n}(t)$ are the eigenvalues of the Hamiltonian \cite{alma9935609690001731, PhysRevA.80.012106, PhysRevLett.95.110407, PhysRevA.76.024304}. Although this condition was shown to not always guarantee the validity of the adiabatic approximation, it still provides a useful framework for analysis \cite{PhysRevLett.95.110407}.

\begin{figure}[b] 
    \includesvg[width=\linewidth]{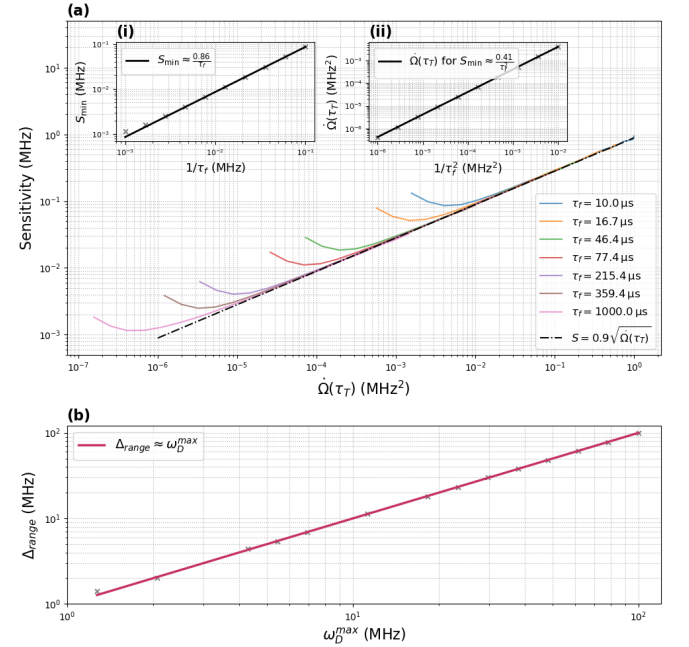}
 \caption{
(a) Simulated ATM fall pulse sensitivity $S$ as a function of the fall gradient $\dot{\Omega}(\tau_T)$ for different fall times $\tau_f$, showing an overall square root scaling $S \propto \sqrt{|\dot{\Omega}(\tau_T)|}$. For each $\tau_f$, a minimum sensitivity is observed, characterized by: (i) an inverse proportionality between fall time and the minimum sensitivity, $S_{\min} \propto \tau_f^{-1}$; and (ii) an inverse square relationship between the final gradient at which the minimum occurs and the fall time, $|\dot{\Omega}(\tau_T)_{S_{\min}}| \propto \tau_f^{-2}$. 
(b) Detuning range as a function of chirped frequency, showing an approximately one-to-one linear relationship, $\Delta_{\mathrm{range}} \approx \omega_D$.
}

    \label{fig:RESULTS}
\end{figure}

Evaluating this condition for the ATM pulse Hamiltonian at $t=\tau_T$ gives a relation between the final gradient $\dot{\Omega}(\tau_T)$ and the detuning $\Delta$, such that
\begin{eqnarray}
    \frac{|\dot{\Omega}(\tau_T)|}{2\Delta^2}\ll1.
\end{eqnarray}
Appendix~\ref{sec:FinalSegment} derives the endpoint value and determines when it is the maximum over the complete fall. For the pulse parameters in Appendix~\ref{sec:Pulse}, the endpoint value is the maximum over the fall for $0<|\Delta|\leq0.08825~\mathrm{MHz}$, or approximately $88~\mathrm{kHz}$. Here, the sensitivity $S:=\left(\max_{\Delta}|dP/d\Delta|\right)^{-1}$ is used as a transition-width metric for the central edge. Smaller $S$ denotes a sharper response. Within this endpoint-controlled regime, the characteristic detuning scale is proportional to $\sqrt{|\dot{\Omega}(\tau_T)|}$. This motivates the square-root dependence, while simulations determine the numerical prefactor shown in Figure \ref{fig:RESULTS}(a):
\begin{eqnarray}
    S \approx 0.9\sqrt{|\dot{\Omega}(\tau_T)|}.
\end{eqnarray}

These simulation results are consistent with the theory. For each fall time $\tau_f$, however, there exists a minimum sensitivity, indicating that if $\tau_f$ is not sufficiently long, the curvature becomes too sharp and adiabaticity is lost too early. As shown in Figure \ref{fig:RESULTS}(a)(i), the minimum sensitivity for a given fall time is given by
\begin{eqnarray}
    S_{\min} \approx \frac{0.85}{\tau_f},
\end{eqnarray}
while the ATM fall gradient required to achieve that minimum sensitivity, as shown in Figure \ref{fig:RESULTS}(a)(ii), is
\begin{eqnarray}
    |\dot{\Omega}(\tau_T)_{S_{\min}}| \approx \frac{0.41}{\tau_f^2}.
\end{eqnarray}
These relationships therefore provide a practical design rule: $\tau_f$ should be chosen to satisfy the minimum adiabatic time required for the tangent profile to reach the desired $\dot{\Omega}(\tau_T)$, and hence the target sensitivity.

\subsection{Optimizing for Range}
As illustrated in Figure \ref{fig:RESULTS}(b), the detuning range has an approximate one-to-one linear relationship to the range of frequencies chirped by the ATM pulse,
\begin{eqnarray}
    \Delta_{\text{range}}\approx\omega^{\max}_D.
\end{eqnarray}
The system's maximum allowable frequency sweep is therefore a key limitation on the range of an ATM experiment.

More precisely, the accessible detuning range $\Delta_{\mathrm{range}}$ is determined by the span between the laboratory qubit frequency and the chirped drive frequency. Provided $\omega_D^{\max}$ includes the worst-case detuning, the ATM pulse can map those detunings into the excited state region. When designing an experiment, one should therefore choose $\omega_D^{\max}$ so that $\Delta_{\mathrm{range}}$ covers the target tracking window. This approximate scaling assumes the chirp remains sufficiently slow to produce the desired response, as discussed in Appendix~\ref{sec:Adiabaticity}.

\section{\label{sec:SLRComparison}Comparison with Shinnar--Le Roux Binary Tracking Pulses}

Binary detuning responses are not unique to the ATM pulse. In magnetic resonance, pulses derived using the Shinnar--Le Roux (SLR) formalism can also be designed to produce sharp transitions between excitation and non-excitation regions \cite{shinnar1989synthesis,grissom2014b1}. These pulses achieve selectivity through Fourier domain filter design and can generate responses resembling the sigmoidal transfer characteristic seen in ATM pulses.

However, the physical mechanisms underlying the two approaches are fundamentally different. SLR pulses rely on precise interference between frequency components of a carefully designed waveform, whereas the ATM pulse derives its selectivity from adiabatic evolution in the instantaneous eigenbasis. As a consequence, SLR pulse performance is highly dependent on accurate control of the applied Rabi amplitude and pulse calibration \cite{grissom2014b1}. In contrast, adiabatic protocols are generally expected to exhibit reduced sensitivity to Rabi-amplitude uncertainty because the state evolution follows the instantaneous Hamiltonian eigenstates rather than a specific resonant trajectory.

To quantify this difference, we compare the ATM pulse with an SLR inversion pulse designed using the standard \texttt{dzrf} implementation. The two pulse types were chosen to have comparable pulse durations ($\sim30~\mu$s), excitation ranges ($\sim0.6$~MHz), and sensitivities ($\sim70$~kHz). Quasistatic amplitude noise was introduced through
\begin{eqnarray}
\Omega \rightarrow \Omega_{\mathrm{nom}}10^{\delta/20},
\end{eqnarray}
where $\delta$ is uniformly distributed over a specified amplitude uncertainty window expressed in dB. The amplitude noise was modeled as quasistatic because slow calibration drift and gain fluctuations are often the dominant source of control amplitude uncertainty in experimental quantum hardware. This approximation is especially relevant for large-scale qubit characterization and frequency tracking applications, where measurements are performed repeatedly across many qubits and slowly varying calibration errors can be treated as constant over the duration of an individual experiment.

\begin{figure}[b]
    \centering
    \includesvg[width=\linewidth]{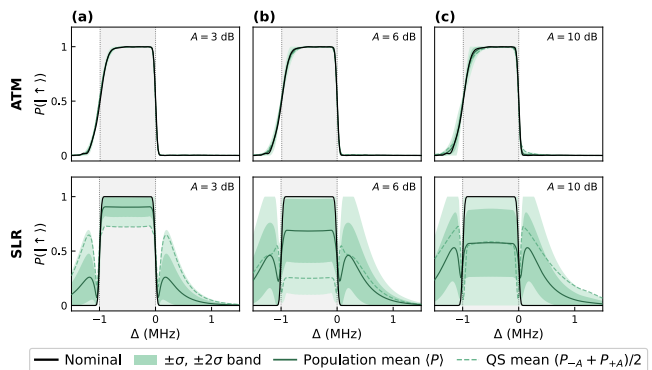}
    \caption{
    Monte Carlo analysis of robustness to quasistatic Rabi-amplitude noise for the ATM pulse (see Appendix~\ref{sec:Pulse} for pulse parameters; top row) and the Shinnar-Le Roux (SLR) inversion pulse (bottom row). Columns (a)--(c) correspond to amplitude noise
    $\Omega \rightarrow \Omega_{\mathrm{nom}}\,10^{\delta/20}$,
    with $\delta \in [-A_{\mathrm{dB}}, +A_{\mathrm{dB}}]$ and
    $A_{\mathrm{dB}} = 3, 6,$ and $10$~dB, respectively.
    The black curve shows the nominal response ($\delta = 0$), the solid green curve shows the averaged population $\langle P \rangle$, and the shaded regions indicate the $\pm\sigma$ and $\pm2\sigma$ regions obtained from $N=512$ iterations. The dashed green curve shows the quasistatic endpoint estimate $(P_{-A}+P_{+A})/2$. Vertical dotted lines indicate the nominal ranges. The SLR response is frequency-shifted so that its nominal transition point coincides with $\Delta=0$, enabling direct comparison of sensitivity degradation under identical amplitude-noise conditions.
    }
    \label{fig:NoiseRobustness}
\end{figure}

Figure~\ref{fig:NoiseRobustness} shows Monte Carlo simulations for amplitude uncertainties of $\pm3$, $\pm6$, and $\pm10$~dB. The ATM response remains largely unchanged across all three noise levels, preserving both its sensitivity edge and excitation range. By contrast, the SLR pulse progressively broadens, shifts, and loses selectivity as the amplitude uncertainty increases. Significant degradation is already visible for $\pm3$~dB amplitude variation and becomes severe by $\pm10$~dB.

These results suggest that the adiabatic mechanism underlying the ATM pulse provides a substantial practical advantage for qubit frequency tracking. Since calibration drift and slow variations in control amplitude are common in experimental quantum hardware, maintaining a stable binary response in the presence of amplitude uncertainty is critical. The robustness demonstrated here allows the ATM pulse to preserve its tracking characteristics without requiring the frequent recalibration that would be necessary for a conventional SLR-based tracker.

\section{\label{sec:NoisePSD}Noise Power Spectrum and Frequency Tracking Performance}

Having established that the ATM pulse maintains its binary detuning response under substantial amplitude uncertainty, we now compare its frequency tracking performance against Ramsey-based feedback. ATM pulses offer two key advantages over Ramsey tracking. First, ATM pulses provide sensitivity over a finite detuning range while Ramsey feedback is inherently periodic. This periodicity can introduce ambiguities, causing the controller to lock onto a detuning offset by an integer multiple of the Ramsey fringe spacing rather than the true detuning. 

\begin{figure}[b] 
    \centering
    \includesvg[width=\linewidth]{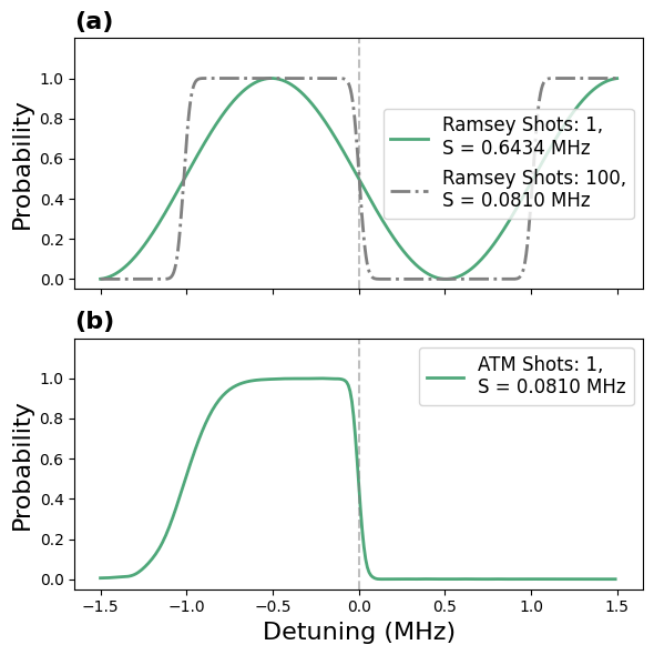}
    \caption{Comparison of Ramsey and ATM detuning sensitivity.
    (a) Transition probability versus detuning for a single-shot Ramsey experiment, yielding a sensitivity of $S = 0.6434$\,MHz over a range $\Delta_{\mathrm{range}} \approx 1$\,MHz, compared with Bayesian estimation using 100 shots, which improves the sensitivity to $S = 0.0810$\,MHz over the same range.
    (b) Transition probability versus detuning for a single $30\,\mu$s ATM pulse (see Appendix \ref{sec:Pulse} for pulse parameters), achieving $S = 0.0810$\,MHz with $\Delta_{\mathrm{range}} \approx 1$\,MHz.}
    \label{fig:Ramsey}
\end{figure}

Second, as shown in Figure~\ref{fig:Ramsey}, Ramsey feedback has significantly worse sensitivity than using ATM and the ratio of gain to sensitivity has been found to be the determining factor in the tracking behavior. The sensitivity of Ramsey feedback can be improved using a Bayesian estimate of multiple Ramsey shots \cite{77qg-p68k}, however, the time for a single measurement is dominated by readout time and thus a single iteration of feedback would take much longer. So while the results at higher frequencies would be more accurate it would limit the highest possible frequency of noise that can be measured. In these simulations, we find that the switching point in tracking behavior occurs near $G = S / 2$ and so we examine the behavior on either side of this point.

\subsubsection{\label{subsec:high}High-gain regime}

Starting with $G \geq S / 2$ we have found the closed-loop noise floor of ATM tracking rises smoothly according to
\begin{eqnarray}
    P_{\text{N}}^{\text{floor}} = \frac{G^2}{f_s}.
\end{eqnarray}
On the other hand, Ramsey tracking, seen in Figure ~\ref{fig:ATMRamseyPSD}(a), develops excess low-frequency noise which is attributed to the previously mentioned periodicity of the detuning response. At gains this large, a random shift in the Larmor frequency between feedback iterations or an incorrect result due to imperfect visibility can easily cause the tracker to cross a fringe boundary and temporarily lock onto an incorrect Ramsey fringe, thus causing an apparent increase in low-frequency noise. We also see with large gain that both trackers lose sensitivity to isolated high-frequency tones, however, ATM retains the ability to detect these tones much better while Ramsey tracking quickly drowns out any trace.

\subsubsection{\label{subsec:low}Low-gain regime}

\begin{figure}[b]
    \includesvg[width=\linewidth]{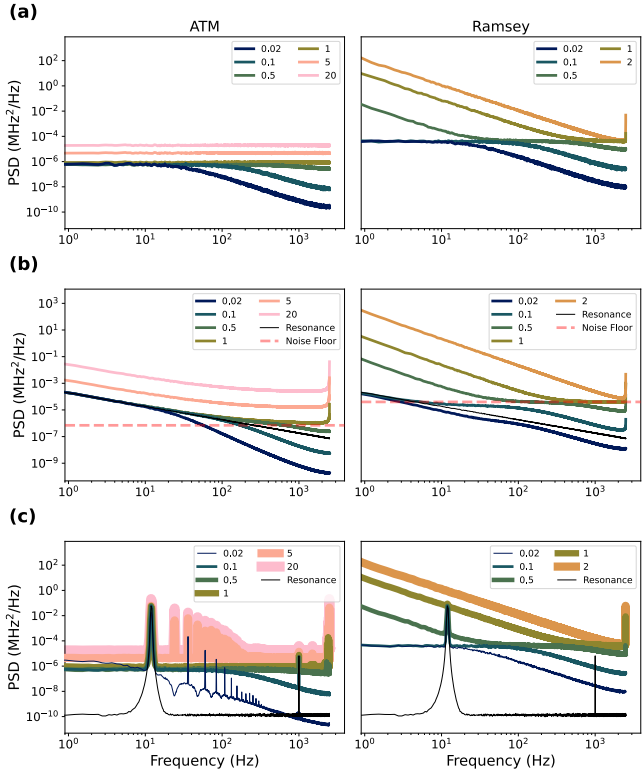}
         \caption{Noise power spectral density of simulated Larmor frequency tracking using ATM pulses and single-shot Ramsey, with each curve corresponding to a different static gain value proportional to the pulse sensitivity region. (a) Noise-free simulated data. ATM exhibits a lower closed-loop noise floor; both trackers show low-pass filtering at low gain, but only Ramsey feedback develops distortion at high gain. (b) Simulated Larmor frequency with $1/f$ noise. With appropriate gain, the lower ATM closed-loop noise floor more accurately reproduces the true PSD at higher frequencies than Ramsey feedback. (c) Simulated Larmor frequency containing two sinusoidal noise tones -- a high-power low-frequency ($0.5$ MHz amplitude, $12$ Hz frequency) and a low power high frequency component ($0.005$ MHz amplitude, $1000$ Hz frequency). ATM resolves the weaker high frequency peak missed by Ramsey tracking, while consistently retaining the stronger low frequency peak even at higher gains.}
    \label{fig:ATMRamseyPSD}
\end{figure}

At $G \leq S / 2$, the idealized controller analyzed in Appendix~\ref{sec:ControllerNoise} has the closed-loop noise floor
\begin{eqnarray}
    P_{\text{N}}^{\text{floor}} = \frac{S^2}{4f_s}.
\end{eqnarray}
In this situation, the applied gain $G$ is small relative to the sensitivity $S$, meaning the per-shot correction is small compared to typical noise fluctuations. This results in the tracker being unable to keep pace with fast changes in detuning; this manifests as a low-pass filter whose cutoff scales with the gain. Both ATM and Ramsey trackers show this behavior at low gain (see Figure ~\ref{fig:ATMRamseyPSD}). Critically, as we have discussed above, Ramsey's usable gain range is bounded well below ATM's to avoid fringe slips, so the Ramsey tracker is restricted to operating in this low gain regime. The low-pass suppression of high-frequency content seen in Figure ~\ref{fig:ATMRamseyPSD}(b,c) is therefore a consequence of operating the feedback controller at low gain \cite{malinowski_spectrum_2017, rower_exploring_nodate, yan_spectroscopy_2012, seedhouse_wavelet_2025}.

Because the frequency controller produces a single-shot binary output, the gain applied at each iteration is critical, with the trade-offs above governing its choice. This limitation can be mitigated using an adaptive-gain feedback controller, for example by casting frequency tracking as a binary search problem \cite{77qg-p68k}.

\section{\label{sec:Conclusion}Conclusions and Outlook}

We have presented the adiabatic tangentially-modulated (ATM) pulse, a single-shot protocol for qubit frequency tracking derived from quantum adiabatic theory. By combining a tangent-shaped amplitude envelope with a piecewise-linear frequency chirp, the ATM pulse produces a sigmoidal, near-binary detuning response whose sensitivity and range are primarily controlled by the fall gradient $\dot{\Omega}(\tau_T)$ and maximum chirped frequency $\omega_D^{\mathrm{max}}$, respectively. The derived scaling relations $S \propto \sqrt{|\dot{\Omega}(\tau_T)|}$ and $\Delta_{\mathrm{range}} \approx \omega_D^{\mathrm{max}}$ provide practical approximate design rules, confirmed by numerical simulation. Importantly, numerical optimization further indicates the minimum achievable sensitivity scales as $S_{\min} \propto \tau_f^{-1}$, where fall time dominates the total ATM pulse length. Unlike binary response pulse designs derived from the Shinnar-Le Roux formalism, the ATM pulse retains its operating characteristics under substantial Rabi amplitude noise. Monte Carlo simulations demonstrate that the excitation range and sensitivity remain largely unchanged even for amplitude variations approaching $\pm10$~dB, whereas comparable SLR pulses exhibit significant degradation. This robustness arises naturally from the adiabatic evolution underlying the ATM protocol and reduces the dependence on precise pulse calibration. Benchmarked against the Ramsey protocol considered here, a single ATM shot achieves sensitivity equivalent to a 100-shot Bayesian Ramsey estimate, enabling accurate tracking of noise components up to two orders of magnitude higher in frequency.

The feedback gain analysis presented here suggests that adaptive gain strategies, informed by the instantaneous noise environment, could improve tracking of both abrupt frequency jumps and weak high-frequency noise tones. In silicon spin qubits, the ATM protocol offers a practical path toward real-time frequency stabilization with reduced susceptibility to the periodic-fringe ambiguities of Ramsey-based feedback. More broadly, the analytical design rules developed here may be applicable to other two-level systems driven by resonant pulses, making the ATM pulse a candidate for frequency tracking in other qubit platforms facing similar noise environments.

\section*{\label{sec:Acknowledgments}Acknowledgments}

We thank Dr Tuomo Tanttu and Andreas Nickl for valuable initial discussions regarding the ATM pulse.

\section*{\label{sec:Contributions}Author Contributions}
C.H.Y. conceived the ATM pulse concept and developed the initial simulation framework. I.O. developed and performed the simulations of the ATM pulse, including the geometric interpretation, detuning parameterization, and pulse optimization. I.O. also developed the theoretical adiabaticity analysis and drafted the manuscript. L.D. investigated the noise power spectrum and frequency tracking performance, contributed to the corresponding manuscript sections, and to additional simulations and theoretical analysis. C.H. verified the adiabaticity calculations, contributed to the drive-following interpretation and manuscript revision, and created the appendix figure. G.A.P.-S. provided guidance on relevant literature and general theoretical analysis, and contributed significantly to the SLR comparison and associated simulations. C.H.Y. supervised the project and contributed to manuscript editing. All authors reviewed and approved the final manuscript.

\section*{\label{sec:Availability}Code and Data Availability}

The code and data supporting the findings of this study will be publicly available in a Git repository.

\clearpage

\appendix
\section{\label{sec:Rotating} Rotating frame conversion}
Given the lab frame Hamiltonian
\begin{eqnarray} 
    \mathcal{H}_{\mathrm{lab}}(t) = \\ \nonumber
    \frac{\hbar}{2}
    \begin{pmatrix}
        \omega_0 & 2\Omega(t)\cos(\omega t+\phi_D(t))\\
        2\Omega(t)\cos(\omega t+\phi_D(t)) & -\omega_0
    \end{pmatrix}
\end{eqnarray}
and a unitary transformation $U=e^{i\sigma_z\frac{\omega t}{2}}$, corresponding to a frame rotating at frequency $\omega$, the rotating-frame Hamiltonian is $\mathcal{H}_{\mathrm{rot}}(t)=U\mathcal{H}_{\mathrm{lab}}(t)U^{\dagger}+i\hbar\dot{U}U^{\dagger}$. Under the rotating-wave approximation, this becomes
\begin{eqnarray}
\label{HRWA}
    \mathcal{H}_{\mathrm{rot}}(t) = \frac{\hbar}{2}
    \begin{pmatrix} 
    \Delta & \Omega(t)e^{-i\phi_D(t)} \\ 
    \Omega(t)e^{i\phi_D(t)} & - \Delta 
    \end{pmatrix}
\end{eqnarray}

where $\Delta = \omega_{0} - \omega$.

\section{\label{sec:Adiabaticity}Adiabaticity of the ATM pulse}

The frame used in the main text rotates at the carrier frequency $\omega$ and provides a geometric picture of the qubit trajectory. To understand adiabatic following throughout the ATM pulse and clarify how its three segments produce the final response, we transform to a frame that follows the drive frequency $\omega + \omega_D(t)$. In this frame, the instantaneous frequency offset and associated avoided crossing appear explicitly. \\

Starting from the Hamiltonian in \eqref{HRWA}, we apply the additional rotation $V(t)=e^{i\sigma_z\phi_D(t)/2}$. The transformed drive-frame Hamiltonian is $\mathcal{H}_{\mathrm{df}}(t)=V(t)\mathcal{H}_{\mathrm{rot}}(t)V^\dagger(t)+i\hbar\dot{V}(t)V^\dagger(t)$, giving
\begin{eqnarray}\label{HDF}
  \mathcal{H}_{\mathrm{df}}(t)=\left[\Delta-\omega_D(t)\right]S_z+\Omega(t)S_x.
\end{eqnarray}
Because $V(t)$ is a rotation about $z$, this basis change leaves the measured qubit populations unchanged. \\

Defining $\delta(t)=\Delta-\omega_D(t)$ as the instantaneous detuning between the qubit and applied drive, the instantaneous eigenenergies of \eqref{HDF} are 
\begin{eqnarray}
  E_\pm(t)=\pm\frac{\hbar}{2}\sqrt{\delta(t)^2+\Omega(t)^2}.
\end{eqnarray}
Without the transverse drive, the two energies cross at $\delta(t)=0$. For $\Omega(t)\neq0$, they remain separated by the finite gap $\hbar|\Omega(t)|$, producing an avoided crossing centered at $\delta(t)=0$. \\

The instantaneous eigenstates of \eqref{HDF} are
\begin{eqnarray}\label{EStates}
    |E_+(t)\rangle
    &=&
    \begin{pmatrix}
        \cos\left(\frac{\theta(t)}{2}\right)\\
        \sin\left(\frac{\theta(t)}{2}\right)
    \end{pmatrix}, \nonumber\\
    |E_-(t)\rangle
    &=&
    \begin{pmatrix}
        -\sin\left(\frac{\theta(t)}{2}\right)\\
        \cos\left(\frac{\theta(t)}{2}\right)
    \end{pmatrix}
\end{eqnarray}
where
\begin{eqnarray}
  \theta(t)=\operatorname{atan2}\left(\Omega(t),\delta(t)\right).
\end{eqnarray}
For a pulse initialized in $|\downarrow\rangle$, the instantaneous eigenbranch selected at the beginning of the pulse is
\begin{eqnarray}
    |E_{\mathrm{sel}}(t)\rangle=
    \begin{cases}
        |E_-(t)\rangle, & \Delta>-\omega_D^{\max},\\
        |E_+(t)\rangle, & \Delta<-\omega_D^{\max}.
    \end{cases}
\end{eqnarray}
At $\Delta=-\omega_D^{\max}$, the initial Hamiltonian is degenerate and the selected eigenbranch is not uniquely defined.\\

\begin{figure}[t]
  \includesvg[width=\linewidth]{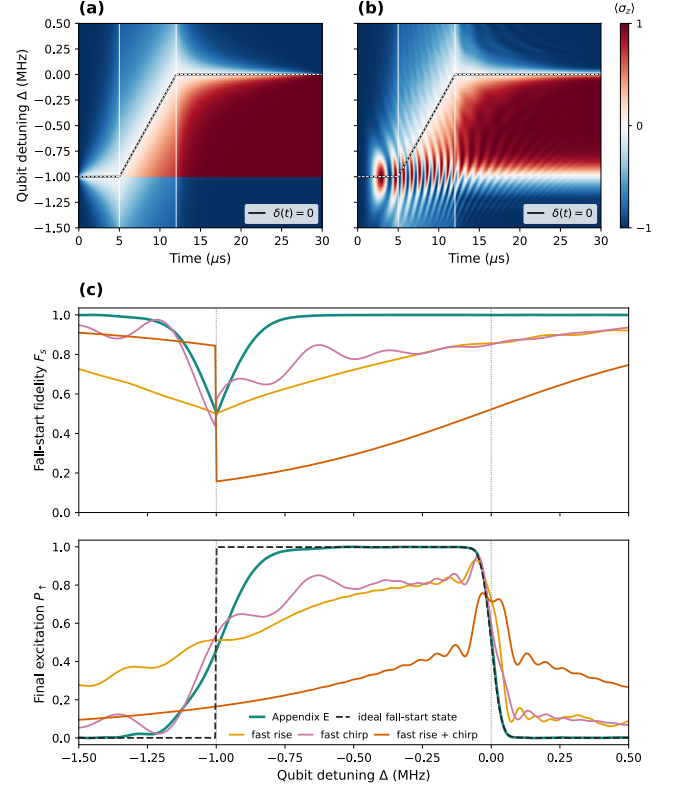}
  \caption{Drive-following dynamics for the ATM pulse with parameters given in Appendix~\ref{sec:Pulse}. (a) The $\langle\sigma_z\rangle$ character of the instantaneous eigenbranch selected by the initial qubit state. (b) The corresponding exact qubit evolution. The black-and-white line marks the avoided-crossing condition $\delta(t)=0$, and the vertical white lines separate the rise, chirp, and fall segments. (c) Fall-start fidelity (top) and final excitation probability (bottom) for the Appendix~\ref{sec:Pulse} pulse, a fast rise with $\tau_r=0.1~\mu\mathrm{s}$ and $\dot{\Omega}(0)=5~\mathrm{MHz}/\mu\mathrm{s}$, a fast chirp with $\tau_c=0.01~\mu\mathrm{s}$, and their combination. The unchanged segments retain the parameters given in Appendix~\ref{sec:Pulse}. The dashed black curve in the lower panel gives the response obtained by initializing the fall exactly in the selected instantaneous eigenstate, thereby isolating the response produced by the fall.}
  \label{fig:AppFig}
\end{figure}

As $\Omega(t)\rightarrow0$, the bare-state character of each eigenbranch is determined by the sign of $\delta(t)$. During the negative chirp, $\delta(t)$ changes from $\Delta+\omega_D^{\max}$ to $\Delta$ and therefore passes through the avoided crossing if and only if
\begin{eqnarray}
  -\omega_D^{\max}<\Delta<0.
\end{eqnarray}
The central chirp has the same finite-gap avoided crossing structure used in adiabatic rapid passage \cite{Malinovsky2001ARP}. Within this interval, following the selected eigenbranch through the crossing reverses its bare-state character. The final fall reduces $\Omega(t)$ to zero, mapping the dressed eigenstate onto the corresponding bare state. Perfect eigenbranch following therefore produces inversion within the swept detuning range and no inversion outside it. At $\Delta=-\omega_D^{\max}$, the initial gap closes, while at $\Delta=0$ the final gap closes. Figure~\ref{fig:AppFig}(a) shows the $\sigma_z$ character of the selected instantaneous eigenbranch throughout the pulse for a range of detunings, while panel (b) shows the exact qubit evolution for the pulse with parameters given in Appendix~\ref{sec:Pulse}. The final values in panel (a) give this ideal binary response, and comparison with panel (b) shows how closely the implemented pulse approaches this limit. The goal of ATM pulse engineering is therefore to keep the qubit state as close as possible to the selected instantaneous eigenbranch throughout the pulse. \\

For a nonzero instantaneous energy gap, a standard local measure of adiabatic following is \cite{PhysRevA.80.012106}
\begin{eqnarray}
      \epsilon_{mn}(t)
      &=&
      \frac{\hbar\left|\langle E_m(t)|\dot{\mathcal{H}}(t)|E_n(t)\rangle\right|}
      {\left|E_m(t)-E_n(t)\right|^2} \nonumber\\
      &=&
      \left|\frac{\hbar\langle E_m(t)|\dot{E}_n(t)\rangle}
      {E_m(t)-E_n(t)}\right|, \qquad m\neq n.
      \label{eq:AdiabaticParameter}
  \end{eqnarray}
$\epsilon_{mn}(t)\ll1$ is an indicator for local adiabatic following. This parameter is diagnostic rather than a transition probability and does not by itself guarantee a small final error \cite{PhysRevLett.95.110407}. \\

Substituting the drive-following eigenstates \eqref{EStates} into \eqref{eq:AdiabaticParameter}, and using $\dot{\delta}(t)=-\dot{\omega}_D(t)$, gives
\begin{eqnarray}\label{eq:DriveAdiabaticParameter}
  \epsilon_{\mathrm{df}}(t)
  =
  \frac{\left|\delta(t)\dot{\Omega}(t)+\Omega(t)\dot{\omega}_D(t)\right|}
  {2\left[\delta(t)^2+\Omega(t)^2\right]^{3/2}}.
\end{eqnarray}
This expression shows how the rise and chirp affect preparation of the selected eigenbranch. During the rise, $\omega_D(t)$ is constant and the increasing amplitude creates a transverse contribution to the energy gap. At $\Delta=-\omega_D^{\max}$, both $\delta(0)$ and $\Omega(0)$ vanish, so the initial energy gap closes. This boundary degeneracy is distinct from the finite-gap avoided crossing traversed during the chirp, and a rapid rise can produce poor eigenbranch preparation nearby. During the constant-amplitude chirp, the adiabatic parameter reaches its maximum at $\delta(t)=0$, provided the avoided crossing lies within the sweep interval. Increasing the chirp rate increases this maximum and the risk of leakage from the selected eigenbranch. The rise and chirp must therefore be sufficiently slow for the state entering the fall to be close to the selected eigenbranch. \\

To quantify this preparation, we define the fall-start fidelity
\begin{eqnarray}\label{eq:FallStartFidelity}
    F_s(\Delta)=\left|\langle E_{\mathrm{sel}}(t_s)|\psi(t_s)\rangle\right|^2, \qquad t_s=\tau_r+\tau_c,
\end{eqnarray}
where $|E_{\mathrm{sel}}(t_s)\rangle$ is the instantaneous eigenstate connected to the initial qubit state. \\

Figure~\ref{fig:AppFig}(c) separates preparation errors accumulated during the rise and chirp from the response produced by the fall. For the Appendix~\ref{sec:Pulse} pulse, $F_s$ remains close to one near $\Delta=0$, and the exact response closely follows the dashed curve obtained with perfect fall-start preparation. The width of the central transition is therefore determined predominantly by the fall. Near the outer edge at $\Delta=-\omega_D^{\max}$, the initial gap becomes small and rise and chirp errors instead control the response; the deliberately faster variants amplify these distortions. \\

At the exact degeneracy, $|E_{\mathrm{sel}}\rangle$ and $F_s$ are undefined, although the propagated state and final population remain continuous. Thus the outer edge limits the capture range, whereas the fall controls the binary decision edge near $\Delta=0$. The fall is analyzed in Appendix~\ref{sec:FinalSegment}.

\section{\label{sec:FinalSegment} Adiabaticity during the final segment}

During the fall, $\dot{\phi}_D(t)=0$, so the drive-following and carrier frames differ only by a time-independent rotation about the $z$-axis. This rotation leaves both the adiabatic parameter and $\langle \sigma_z \rangle$ unchanged. The following analysis therefore applies directly to the above- and below-equator trajectories in Figure~\ref{fig:BLOCH}. Equation~\eqref{eq:DriveAdiabaticParameter} reduces to
\begin{eqnarray}
     \epsilon_{\rm fall}(t) = 
    \frac{1}{2}\frac{\lvert \Delta \dot{\Omega}(t) \rvert}{(\Delta^2+\Omega(t)^2)^\frac{3}{2}}.
\end{eqnarray}
$\Omega(\tau_T)=0$, so for $\Delta \neq 0$,
\begin{eqnarray}
     \epsilon_{\rm fall}(\tau_T) =  
    \frac{|\dot{\Omega}(\tau_T)|}{2\Delta^2}.
\end{eqnarray}
Although exact at the endpoint, this value need not be the maximum over the complete fall. We therefore maximize $\epsilon_{\mathrm{fall}}(t)$ for the tangent-shaped envelope. \\

For $\tau_r + \tau_c \leq t \leq \tau_T$, we have
\begin{eqnarray}
    \Omega(t) =  -\frac{|\dot{\Omega}(\tau_T)|}{\beta_f} \tan\left[\beta_f (t-\tau_T)\right], \\
    \dot{\Omega}(t) = -|\dot{\Omega}(\tau_T)| \sec^2\left[\beta_f (t-\tau_T)\right], \\
    \tan(\theta(t)) = \frac{\Omega(t)}{\Delta}.
\end{eqnarray}
The final ratio applies for $\Delta \neq 0$. Throughout the implemented fall, $0 \leq \beta_f(\tau_T-t) < \frac{\pi}{2}$ so that the tangent envelope remains finite and monotonic. To calculate the maximum of $\epsilon_{\rm fall}(t)$, define

\begin{eqnarray}
  y:=\tan^2\left[\beta_f(t-\tau_T)\right],\qquad 
  \kappa:=\frac{|\dot{\Omega}(\tau_T)|}{\beta_f |\Delta|}.
\end{eqnarray}
Since
\begin{eqnarray}
  |\dot{\Omega}(t)|=|\dot{\Omega}(\tau_T)|(1+y)
\end{eqnarray}
and
\begin{eqnarray}
  \Omega(t)^2=\frac{|\dot{\Omega}(\tau_T)|^2}{\beta_f^2}y,
\end{eqnarray}
the adiabatic parameter becomes
\begin{eqnarray}
  \epsilon_{\rm fall}(y) = \frac{|\dot{\Omega}(\tau_T)|}{2|\Delta|^2} \frac{1+y}{\left(1+\kappa^2y\right)^{3/2}}.
\end{eqnarray}
Differentiating with respect to $y$ gives
\begin{eqnarray}
  \frac{d\epsilon_{\rm fall}}{dy} =
  \frac{|\dot{\Omega}(\tau_T)|}{4|\Delta|^2} \frac{2-3\kappa^2-\kappa^2y}{\left(1+\kappa^2y\right)^{5/2}}.
\end{eqnarray}
The numerator decreases monotonically with $y$. Therefore, the endpoint $y=0$, corresponding to $t=\tau_T$, is the maximum if and only if
\begin{eqnarray}
  2-3\kappa^2\leq0,
\end{eqnarray}
or equivalently
\begin{eqnarray}
  |\Delta|\leq\sqrt{\frac{3}{2}}\,\frac{|\dot{\Omega}(\tau_T)|}{\beta_f}.
\end{eqnarray}
Within this detuning interval,
\begin{eqnarray}
  \epsilon_{\rm fall}^{\max}=\epsilon_{\rm fall}(\tau_T)=\frac{|\dot{\Omega}(\tau_T)|}{2\Delta^2}.
\end{eqnarray}

Outside this interval, the unconstrained stationary point has Rabi amplitude
\begin{eqnarray}
  \Omega_* := \Omega(t_*) = \sqrt{2\Delta^2 - 3\left(\frac{|\dot{\Omega}(\tau_T)|}{\beta_f} \right)^2}.
\end{eqnarray}
Defining the amplitude at the beginning of the fall as
\begin{eqnarray}
  \Omega_{\rm start}:=\Omega(\tau_T-\tau_f)=\frac{|\dot{\Omega}(\tau_T)|}{\beta_f}\tan(\beta_f\tau_f),
\end{eqnarray}
the stationary point lies within the implemented fall only when $\Omega_*\leq\Omega_{\rm start}$. If this is not satisfied, the maximum occurs at the beginning of the fall. The corresponding maximum value is obtained by evaluating
\begin{eqnarray}
  \epsilon_{\rm fall}(\Omega) =
  \frac{|\Delta\dot{\Omega}(\tau_T)| \left[1+\left(\frac{\beta_f\Omega}{|\dot{\Omega}(\tau_T)|}\right)^2\right]}
  {2\left(\Delta^2+\Omega^2\right)^{3/2}}
\end{eqnarray}
at $\Omega=\Omega_*$ or $\Omega=\Omega_{\rm start}$, respectively.

At the beginning of the final segment, continuity with the constant-amplitude segment gives
\begin{eqnarray}
  \Omega_{\rm start}=\Omega^{\max},
\end{eqnarray}
and therefore
\begin{eqnarray}
  |\dot{\Omega}(\tau_T)| = \frac{\Omega^{\max}\beta_f}{\tan(\beta_f\tau_f)}.
\end{eqnarray}
This analysis applies to the complete tangent-shaped fall. It does not apply at $\Delta=0$, where the final energy gap closes. 

\section{\label{sec:ControllerNoise} Intrinsic tracker noise}

To find the noise intrinsic to the ATM tracker itself we look at a situation where there is no noise in the Larmor frequency we are tracking. Alongside this we will assume we start on resonance with a gain equal to half the sensitivity region. Given this, there are only 3 possible values we could be detuned from the true resonance by: either we are on resonance ($\Delta$ = 0), or we are above/below resonance by half the sensitivity ($\Delta = \pm S / 2$).

Let $x_n$ denote the detuning of the controller's frequency estimate from the true Larmor after feedback iteration $n$. Let the PSD be 
\begin{align}
    S_{xx}(f_k) &= \frac{2f_s}{N}\left|\frac{1}{f_s}\sum_{n=0}^{N - 1} x_n e^{\frac{i2\pi f_k n}{f_s}}\right|^2 \\
    &= \frac{2}{f_sN}\sum_{n=0}^{N - 1}\sum_{m=0}^{N - 1} x_nx_m e^{\frac{i2\pi k(n - m)}{N}},
\end{align}
where $f_k = k f_s/N$ is the frequency of the k-th discrete Fourier-transform bin. For positive, non-Nyquist frequencies, the factor of 2 accounts for the omitted negative-frequency contribution in the one-sided PSD. One recognizes that when the controller is such that $\Delta = \pm S / 2$ then it will always go to $\Delta = 0 $ and from there it will be a $50\%$ chance to go either up or down. As shown below, this sequence has zero autocorrelation at nonzero lags and therefore has a flat expected PSD (i.e. white noise). Computing the expectation value, one finds 
\begin{align}
    E\left[S_{xx}(f_k)\right] &= \frac{2}{f_sN}\sum_{n=0}^{N - 1}\sum_{m=0}^{N - 1} E\left[x_nx_m\right] e^{\frac{i2\pi k(n - m)}{N}} \\
    &= \frac{2}{f_sN}\sum_{n=0}^{N - 1} E\left[x_n^2\right]
\end{align}
since, for $n\neq m$, $E[x_n x_m]=0$: if either sample is at $\Delta=0$, their product vanishes, while for two nonzero samples the signs are selected independently with equal probability. Finally, the controller alternates between $x_n=0$ and $x_n = \pm S/2$, so that half the samples have $E[x_n^2]=0$ and half have $E[x_n^2] = S^2/4$, we get 
\begin{eqnarray}
    E\left[S_{xx}(f_k)\right] = \frac{S^2}{4f_s}
\end{eqnarray}

\section{\label{sec:Pulse} Pulse parameters}

Unless otherwise stated, the simulations presented in Figures~\ref{fig:BLOCH}, \ref{fig:EvolvedProb}, \ref{fig:Ramsey} and \ref{fig:AppFig} use $\tau_r = 5~\mu\mathrm{s}$, $\tau_c = 7~\mu\mathrm{s}$, $\tau_f = 18~\mu\mathrm{s}$, and $\tau_T = 30~\mu\mathrm{s}$. The drive-frequency offset is swept from $-\omega_D^{\max}$ to zero, with $\omega_D^{\max} = 1~\mathrm{MHz}$ and $\Omega^{\max}=1~\mathrm{MHz}$. Figure~\ref{fig:NoiseRobustness} instead uses $\omega_D^{\max}=0.6~\mathrm{MHz}$, with all other baseline pulse parameters unchanged. The pulse-shape parameters are $\beta_r=0.233112 ~\mu\mathrm{s}^{-1} $ and $\beta_f=0.0835791~\mu\mathrm{s}^{-1}$, obtained by enforcing continuity with endpoint slopes $\dot{\Omega}(0)=0.1~\mathrm{MHz}/\mu\mathrm{s}$ and $|\dot{\Omega}(\tau_T)|=0.006~\mathrm{MHz}/\mu\mathrm{s}$, respectively. \\

For these parameters, we can evaluate the endpoint condition from Appendix~\ref{sec:FinalSegment} to be
\begin{eqnarray}
    0<|\Delta|\leq \sqrt{\frac{3}{2}} \frac{|\dot{\Omega}(\tau_T)|}{\beta_f} =0.08825~\mathrm{MHz}.
\end{eqnarray}
Thus, the endpoint is the maximum for detunings up to approximately $88.25~\mathrm{kHz}$ in magnitude.

\bibliography{bibliography}

\end{document}